\documentclass[12pt]{article}
\usepackage{latexsym,amssymb,amsmath,amscd,amsfonts,epsfig,psfrag,graphicx}
\usepackage{lineno}

\newtheorem{thm}{Theorem}[section]
\newtheorem{col}[thm]{Corollary}

\def\pf{\noindent{\it Proof.\;\;\:}}
\def\qed{\nopagebreak\hfill{\rule{4pt}{7pt}}
\medbreak}

\title{\bf Complexity of the conditional\\ colorability of
graphs\footnote{Supported by NSFC, PCSIRT and the ``973" program.}}

\author{
\small Xueliang Li, Xiangmei Yao and Wenli Zhou\\
\small Center for Combinatorics and LPMC-TJKLC, Nankai University\\
\small Tianjin 300071, P.R. China. Email: lxl@nankai.edu.cn\\
}
\date{}

\begin{document}

\maketitle

\begin{abstract}
For an integer $r>0$, a conditional $(k,r)$-coloring of a graph $G$
is a proper $k$-coloring of the vertices of $G$ such that every
vertex $v$ of degree $d(v)$ in $G$ is adjacent to vertices with at
least $min\{r, d(v)\}$ different colors. The smallest integer $k$
for which a graph $G$ has a conditional $(k,r)$-coloring is called
the $r$th order conditional chromatic number, denoted by
$\chi_r(G)$. It is easy to see that the conditional coloring is a
generalization of the traditional vertex coloring for which $r=1$.
In this paper, we consider the complexity of the conditional
colorings of graphs. The main result is that the conditional
$(3,2)$-colorability is $NP$-complete for triangle-free graphs with
maximum degree at most 3, which is different from the old result
that the traditional 3-colorability is polynomial solvable for
graphs with maximum degree at most 3. This also implies that it is
$NP$-complete to determine if a graph of maximum degree 3 is
$(3,2)$- or $(4,2)$-colorable. Also we have proved that some old
complexity results for traditional colorings still hold for the conditional colorings.\\[2mm]
{\bf Keywords.} vertex coloring, conditional coloring, (conditional)
chromatic number, $NP$-complete\\[2mm]
{\bf AMS Subject Classification 2000.} 05C15, 05C85, 68Q25

\end{abstract}

\section{Introduction}

We follow the terminology and notations of \cite{Bondy} and, without
loss of generality, consider simple connected graphs only.
$\delta(G)$ and $\Delta(G)$ denote the minimum degree and maximum
degree of a graph $G$, respectively. For a vertex $v\in V(G)$, the
{\it neighborhood} of $v$ in $G$ is $N_G(v)=\{u\in V(G): u$ is
adjacent to $v$ in $G\}$, and the degree of $v$ is $d(v)=|N_G(v)|$.
Vertices in $N_G(v)$ are called {\it neighbors of} $v$. $P_n$
denotes the path of $n$ vertices. An edge $e$ is said to be
$subdivided$ when it is deleted and replaced by a path of length two
connecting its ends, the internal vertex of this path is a new
vertex.

For an integer $k>0$. A {\it proper} $k$-coloring of a graph $G$ is
a surjective map $c:\ V(G)\rightarrow\{1,2,\ldots,k\}$ such that if
$u, v$ are adjacent vertices in $G$, then $c(u)\neq c(v)$. The
smallest $k$ such that $G$ has a proper $k$-coloring is the {\it
chromatic number} of $G$, denoted by $\chi (G)$.

In the following we will consider a generalization of the
traditional coloring. For integers $k>0$ and $r>0$, a {\it proper}
$(k,r)$-coloring of a graph $G$ is a surjective map $c:
V(G)\rightarrow \{1,2,\ldots,k\}$ such that both of the following
two conditions hold:
\begin{itemize}
\item[(C1)] if $u,v \in V(G)$ are adjacent vertices in $G$, then
$c(u)\neq c(v)$; and \item[(C2)] for any $v\in V(G)$,
$|c(N_G(v))|\geq min\{d(v), r\}$, where and in what follows,
$c(S)=\{c(u)|u\in S$ for a set $S\subseteq V(G)\}$.
\end{itemize}

For a given integer $r>0$, the smallest integer $k>0$ such that
$G$ has a proper $(k,r)$-coloring is the ($r$th order)
$conditional\ chromatic\ number$ of $G$, denoted by $\chi_r(G)$.

By the definition of $\chi_r(G)$, it follows immediately that $\chi
(G)=\chi_1(G)$, and so $\chi_r(G)$ is a generalization of the
traditional graph coloring. The conditional chromatic number has
very different behavior from the traditional chromatic number. For
example, when $r=2$, from \cite{Lai 3} we know that for many graphs
$G$, $\chi_2(G-v)>\chi_2(G)$ for at least one vertex $v$ of $G$, and
there are graphs $G$ for which $\chi_r(G)-\chi(G)$ may be very
large.

From \cite{Lai 2} we know that if $\Delta (G)\leq 2$, for any $r$ we
can easily have an algorithm of polynomial time to give the graph
$G$ a $(k,r)$-coloring. In \cite{Lai 1}, Lai, Montgomery and Poon
got an upper bound of $\chi_2(G)$ that if $\Delta(G)\geq 3$, then
$\chi_2(G)\leq \Delta(G)+1$. The proof is very long compared with
the proof of a similar result for the traditional coloring. In
\cite{Lai 2} and \cite{Lai 3}, Lai, Lin, Montgomery, Shui and Fan
got many new and interesting results on the conditional coloring. In
the present paper, we are going to investigate the complexity of
deciding if a graph is $(k,r)$-colorable. We first give a simple
proof that for any $k\geq 3$ and $r\geq 2$ it is $NP$-complete to
check if a graph is $(k,r)$-colorable. Then we give the main theorem
in the paper that the conditional $(3,2)$-colorability is
$NP$-complete for triangle-free graphs with maximum degree at most
3, which is different from the old result that the traditional
3-colorability is polynomial solvable for graphs with maximum degree
at most 3. At last we show that the $(3,2)$-colorability is also
$NP$-complete for some special classes of graphs, planar graphs,
hamiltonian graphs and so on.

\section{The complexity of the conditional colorings}

In this section, we shall analyze the complexity of the
$(k,r)$-colorability of graphs. We refer to \cite{Garey1} for
terminology, notations and basic results on complexity not given
here.

If a connected graph $G$ has only one vertex, then $\chi_r(G)=1$; if
a connected graph $G$ has only two vertices, then $\chi_r(G)=2$. For
the other connected graphs $G$, we have $\chi_r(G)\geq 3$ for $r\geq
2$. But the following theorem show that for any $2\leq r<k$ the
$(k,r)$-Col is {\it NP}-complete.

The ($k,r$)-colorable problem, denoted by ($k,r$)-Col, is defined as
follows:

{\bf Input}: A graph $G=(V,E)$ and two integers $k>r\geq 2$.

{\bf Question}: Can one assigns each vertex a color, so that only
$k$ colors are used and the two conditions C1 and C2 are satisfied?
i.e., Is $\chi_r(G)\leq k$?

\begin{thm}\label{thm1}
For every fixed $(k,r)$, $2\leq r<k$, $(k,r)$-Col is NP-complete.
\end{thm}

\pf First, it is easy to see that the problem ($k,r$)-Col is in
$NP$.

Second, it is known that the traditional $k$-colorable problem is
$NP$-complete. So, to show the $NP$-completeness, it is sufficient
to reduce the traditional $k$-colorable problem to the ($k,r$)-Col.
We want to relate any instance $G$ of the $k$-colorable problem to a
graph $G'$, such that $G$ is $k$-colorable if and only if $G'$ is
($k,r$)-colorable.

For each vertex $v$ in $V(G)$, we add a new complete graph $K_r$
and add new edges such that $v$ and $K_r$ form a complete graph of
order $r+1$. The resultant graph is denoted by $G'$. So, $G'$ has
$(r+1)|V(G)|$ vertices, and every vertex in $G'$ is contained in a
$K_{r+1}$. It is easy to see that $G$ is $k$-colorable if and only
if $G'$ is $(k,r)$-colorable. \qed

It is known that in traditional colorings, all graphs with maximum
degree 3 are 3-colorable except for $K_4$ (by Brook's theorem). So
the 3-colorable problem is trivial for this class of graphs. But the
next theorem tells us that the problem $(3,2)$-Col remains
$NP$-complete for triangle-free graphs with maximum degree 3.

\begin{thm}\label{thm4}
The problem $(3,2)$-Col remains $NP$-complete for triangle-free
graphs with maximum degree at most 3.
\end{thm}
\pf First, the problem $(3,2)$-Col for triangle-free graphs with
maximum degree at most 3 is obviously in $NP$.

We want to modify the method given in \cite{Garey} to prove the
$NP$-completeness. We reduce the 3-SAT problem to the problem
$(3,2)$-Col for graphs with maximum degree at most 3. We want to
relate any instance $I$ of the 3-SAT problem to a graph $G$ with
$\Delta(G)\leq 3$, such that $I$ is satisfiable if and only if $G$
is $(3,2)$-colorable. Let the set of literals, of the input $I$ to
the 3-SAT problem, be
$\{x_1,x_2,\ldots,x_n,\bar{x}_1,\bar{x}_2,\ldots,\bar{x}_n\}$ and
the clauses be $C_1,C_2,\ldots,C_m$.

\begin{figure}[h]
\begin{center}
\psfrag{u}{$u_1$} \psfrag{v}{$u_2$} \psfrag{w}{$u_3$}
\psfrag{x}{$v$} \psfrag{G}{$G$} \psfrag{H}{$H$}
\includegraphics[width=12cm]{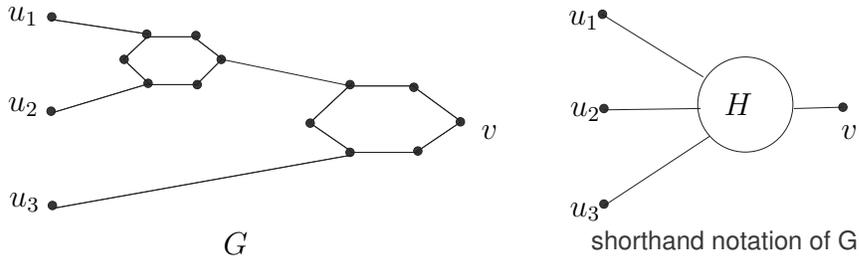}
\caption{The building-block}\label{complexity1}
\end{center}
\end{figure}

The graph $G(V,E)$ with $\Delta(G)\leq 3$ is defined as follows:

First, for each clause $C_i$ ($1\leq i\leq m$) we construct the
first kind of building-block $H_i$ (see Figure \ref{complexity1}).
The second graph in Figure \ref{complexity1} is the shorthand
notation of $G$.

Now, it is easy to check that the graph constructed in Figure
\ref{complexity1} has the following two properties:

\begin{itemize}
\item[1.] If we use colors \{0,1,2\}, and $u_1,u_2,u_3$ are all
colored by 0, then in every $(3,2)$-coloring, $v$ is forced to be
colored by 0; \item[2.] If the three vertices $u_1,u_2,u_3$ are
colored only by 0 or 1 (the other vertices in $G$ can also be
colored by 0, 1 or 2), and not all the three vertices are colored
by 0, then there is a $(3,2)$-coloring such that $v$ can be
colored by 1.
\end{itemize}

From the above properties we know that if $v$ is not colored by 0,
then one of $u_1,u_2,u_3$ must be colored by 1, which means that if
1 represents `true', 0 represents `false', and the clause $C=u_1\vee
u_2\vee u_3$ must be satisfied.

Second, we construct two paths $P_{6n-1}$ ($P_{6n-1}=a_1a_2\cdots
a_{6n-1}$) and $P_{3m-2}$ ($P_{3m-2}=b_1b_2\cdots b_{3m-2}$).

\begin{figure}[h]
\begin{center}
\psfrag{a1}{$a_1$} \psfrag{a2}{$a_4$} \psfrag{a7}{$a_{6n-2}$}
\psfrag{a8}{$a_{6n-1}$} \psfrag{b1}{$b_1$}\psfrag{b2}{$b_4$}
\psfrag{b3}{$b_{3m-2}$} \psfrag{H1}{$H_1$} \psfrag{H2}{$H_2$}
\psfrag{Hm}{$H_m$} \psfrag{G}{$G$} \psfrag{v1}{$v_1$}
\psfrag{v2}{$v_2$} \psfrag{vm}{$v_m$} \psfrag{B1}{$B_1$}
\psfrag{B2}{$B_2$} \psfrag{B3}{$B_3$} \psfrag{B4}{$B_4$}
\psfrag{Bn}{$B_n$} \psfrag{Bi}{$B_i$}
\includegraphics[width=10cm]{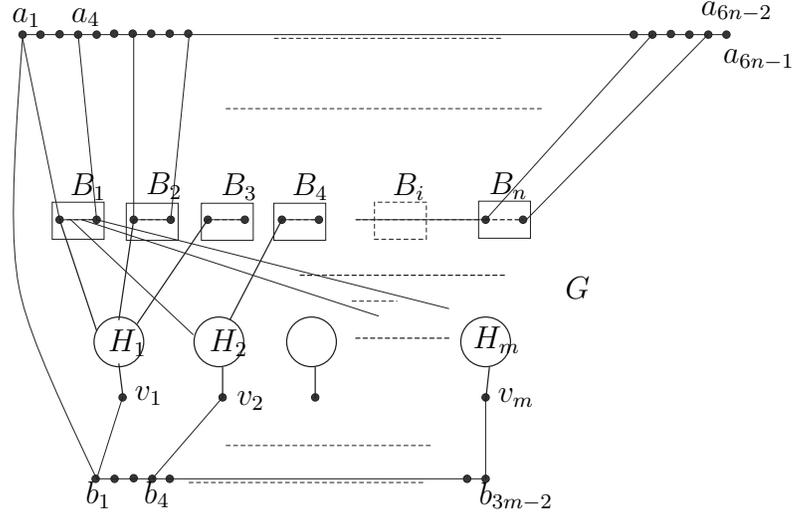}
\caption{The entire construction}\label{complexity2}
\end{center}
\end{figure}

Third, for each pair $x_i$ and $\bar{x}_i$ ($1\leq i\leq n$) we
construct the second kind of building-block $B_i$ (it is
represented by a rectangle in Figure \ref{complexity2}). The
second kind of building-block $B_i$ ($1\leq i\leq n$) is
constructed as follows:

\begin{itemize}
\item[1.] Let $t_i$ be the number of clauses which contain $x_i$, and let
$\bar{t}_i$ be the number of clauses which contain $\bar{x}_i$;
\item[2.] We construct a path $P_{x_i}$ of $3t_i-2$ vertices
corresponding to the vertex $x_i$, and construct another path
$P_{\bar{x}_i}$ of $3\bar{t}_i+2$ vertices corresponding to the
vertex $\bar{x}_i$;
\item[3.] Let $x_{i_j}$ be the $(3j-2)$th $(1\leq j\leq t_i)$ vertex in
the path $P_{x_i}$, and let $\bar{x}_{i_j}$ be the $(3j-2)$th
$(1\leq j\leq \bar{t}_i+1)$ vertex in the path $P_{\bar{x}_i}$;
\item[4.] Join $x_{i_{t_i}}$ with an edge to $\bar{x}_{i_1}$.
\end{itemize}

Finally, join $x_{i_1}$ with an edge to $a_{6i-5}$ and join
$\bar{x}_{i_1}$ with an edge to $a_{6i-2}$ ($1\leq i\leq n$). And
each $v_i$ is joined with an edge to $b_{3i-2}$ ($1\leq i\leq m$).
The vertex $a_1$ is joined with an edge to the vertex $b_1$. Each
$x_{i_j}$ ($1\leq j\leq t_i$) or $\bar{x}_{i_j}$ ($2\leq j\leq
t_i+1$) joins with an edge to some $H_l$ ($1\leq l\leq m$) which
represents the clause that contains the $j$th $x_i$ or the $(j-1)$th
$\bar{x}_i$. The final resultant graph is shown in Figure
\ref{complexity2}. It is easy to see that the maximum degree of the
graph is at most 3 and the graph is triangle-free.

The graphs $B_i$ ($1\leq i\leq n$) in the final resultant graph
(Figure \ref{complexity2}) have the following properties:

\begin{itemize}
\item[1.] If $G$ is $(3,2)$-colorable, the vertices $x_{i_j}$
($1\leq j\leq t_i$) must be colored by the same color, and the
vertices $\bar{x}_{i_j}$ ($1\leq j\leq \bar{t}_i+1$) must also be
colored by the same color. But \item[2.] The color of the vertices
$x_{i_j}$ ($1\leq j\leq t_i$) must be colored by different color
from the color of the vertices $\bar{x}_{i_j}$ ($1\leq j\leq
\bar{t}_i+1$).
\end{itemize}

The two paths $P_{6n-1}$ and $P_{3m-2}$ in the final resultant graph
have the following properties:

\begin{itemize}
\item[1.] If $G$ is $(3,2)$-colorable, the vertices $a_{3i-2}$
($1\leq j\leq 2n$) must be colored the same color, and the
vertices $b_{3i-2}$ ($1\leq i\leq m$) must also be colored by the
same color. But \item[2.] The color of the vertices $a_{3i-2}$
($1\leq i\leq 2n$) must be colored by different color from the
color of the vertices $b_{3i-2}$ ($1\leq i\leq m$).
\end{itemize}

Now if $I$ is satisfiable, we give a proper $(3,2)$-coloring of
the graph in Figure \ref{complexity2} as follows: Let
$c(x_{i_j})=1$ ($1\leq j\leq t_i$) if $x_i$ is true, and of course
let $c(\bar{x}_{i_j})=0$ ($1\leq j\leq \bar{t}_i+1$); let
$c(x_{i_j})=0$ ($1\leq j\leq t_i$) if $x_i$ is false, and of
course let $c(\bar{x}_{i_j})=1$ ($1\leq j\leq \bar{t}_i+1$). The
rest vertices in $B_i$ can be colored easily to satisfy the
conditions C1 and C2. By the properties given in the first step,
we know that $v_i$ ($i=1,\ldots,m$) can be colored by 1 since each
clause is satisfied. Let $c(a_i)=2$ ($i=1,4,7,\ldots,6n-2$),
$c(a_i)=1$ ($i=2,5,8,\ldots,6n-1$), $c(a_i)=0$
($i=3,6,9,\ldots,6n-3$); let $c(b_i)=0$ ($i=1,4,7,\ldots,3m-2$),
$c(b_i)=1$ ($i=2,5,8,\ldots,3m-3$), $c(b_i)=2$
($i=3,6,9,\ldots,3m-4$). Then it is easy to check that $c$ is a
proper $(3,2)$-coloring.

Conversely, if $G$ is $(3,2)$-colorable, there is a proper
$(3,2)$-coloring $c$. Without loss of generality, suppose $c(a_1)=2$
and $c(b_1)=0$. Then, from the properties of the two paths we give
above, all the vertices $a_{3i-2}$ ($1\leq i\leq 2n$) are colored by
2, while all the vertices $b_{3i-2}$ ($1\leq i\leq m$) are colored
by 0. By the properties of the graph $B_i$ described above,
$c(x_{i_j})$ ($1\leq j\leq t_i$) and $c(\bar{x}_{i_1})$ ($1\leq
j\leq \bar{t}_i+1$) are colored by 1 or 0, and $c(v_i)$ cannot be
colored by 0. Then, let $x_i$ be true if $c(x_{i_j})=1$, and let
$x_i$ be false if $c(x_{i_j})=0$. Since $c(v_i)$ cannot be colored
by 0, each $C_i$ is satisfiable, and thus $I$ is satisfiable. \qed

From \cite{Lai 2} we know that if $\Delta (G)=1$ or 2, $\chi _2(G)$
can be determined in polynomial time. From \cite{Lai 1} we know that
if $\Delta (G)=3$, then $\chi_2 (G)=3$ or 4. So by Theorem
\ref{thm4} we can get the following result.
\begin{col}
When $\Delta(G)=3$, it is $NP$-complete to determine whether
$\chi_2(G)=3$ or $\chi_2(G)=4$.
\end{col}

Next we will consider the other special classes of graphs,
hamiltonian graphs, planar graphs, claw-free graphs.

\begin{thm}\label{thm2}
The problem $(3,2)$-Col is NP-complete when restricted to
hamiltonian graphs with $\Delta (G)\leq 6$.
\end{thm}

\pf  A known result is that to determine whether a hamiltonian graph
with maximum degree at most 4 is 3-colorable is $NP$-complete. Now,
given a hamiltonian graph $G$ with $V(G)=\{v_1,v_2,\ldots,v_n\}$
and, without loss of generality, $v_1v_2\ldots v_nv_1$ is a
hamiltonian cycle of $G$. We construct a new hamiltonian graph $G'$
as follows: For each $v_i$ we add two new vertices $x_{i1}$ and
$x_{i2}$ and three new edges $v_ix_{i1}$, $x_{i1}x_{i2}$ and
$x_{i2}v_i$ (a triangle). Then, add new edges $x_{12}x_{21}$,
$x_{32}x_{41}$, $x_{52}x_{61}$, \ldots, $x_{(n-1)2}x_{n1}$ for $n$
even; add new edges $x_{12}x_{21}$, $x_{32}x_{41}$, $x_{52}x_{61}$,
\ldots, $x_{(n-2)2}x_{(n-1)1}$ and add a new vertex $u$ and three
edges $x_{n1}u$, $x_{n2}u$ and $x_{(n-1)2}u$ for $n$ odd.

It is easy to see that $G'$ is also a hamiltonian graph with $\Delta
(G)\leq 6$. First, if $G'$ is $(3,2)$-colorable, then restrict a
proper $(3,2)$-coloring to the vertices $v_1,v_2,\ldots,v_n$, it is
a proper coloring for $G$. Second, if $G$ is 3-colorable, then the
vertices $v_1,v_2,\ldots,v_n$ in $G'$ are colored by the same color
as they are colored in $G$, and since there are 3 colors, the rest
vertices of $G'$ can be colored properly, and so $G'$ is
3-colorable. Since every vertex in $G'$ is contained in a triangle,
$G'$ is 3-colorable means $G'$ is $(3,2)$-colorable. Then $G$ is
3-colorable if and only if $G'$ is $(3,2)$-colorable. Then we get
the result. \qed

Now we consider planar graphs. From \cite{Bonsma} we know that the
3-colorable problem for planar hamiltonian graphs is $NP$-complete.
Then we have the following theorem.

\begin{thm}
The problem $(3,2)$-Col is NP-complete for planar hamiltonian
graphs.
\end{thm}
\pf Given a planar hamiltonian graph $G$ with
$V(G)=\{v_1,v_2,\ldots,v_n\}$ and, without loss of generality,
$v_1v_2\ldots v_nv_1$ is a hamiltonian cycle $C_n$ of $G$. For any
edge $v_iv_{i+1}$ of $C_n$, we do the local transformation to get a
new graph $G'$ as follows: For each edge $v_iv_{i+1}$ in $C_n$, we
add 4 new vertices $x_{i1}$, $x_{i2}$, $y_{(i+1)1}$, $y_{(i+1)2}$
and 7 new edges $x_{i1}v_i$, $x_{i2}v_i$, $x_{i1}x_{i2}$,
$y_{(i+1)1}v_{i+1}$, $y_{(i+1)2}v_{i+1}$, $y_{(i+1)1}y_{(i+1)2}$,
$x_{i2}y_{(i+1)1}$, two triangles with an edge connecting them. Then
$G'$ has $5|V(G)|$ vertices, and every vertex is in a triangle, and
moreover, each ``two triangles with an edge joining them" can be
drawn in the local space of $v_iv_{i+1}$ without crossing the
boundary of any face, so that the new graph $G'$ remains planar. It
is easy to see that is is also hamiltonian. By the same reason as in
Theorem \ref{thm2}, it is easy to see that $G$ is 3-colorable if and
only if $G'$ is $(3,2)$-colorable. The proof is complete. \qed

For claw-free graph $G$, we can similarly add vertices and edges to
make every vertex $v\in V(G)$ in a triangle, and show that
$(3,2)$-Col is NP-complete for claw-free graphs.

To conclude the paper, we point out that there are polynomial
algorithms to solve the $k,r$)-Col problem for some special classes
of graphs. From the proof of \cite{Lai 1}, one can design a
polynomial algorithm to color the graph $G$ by $\Delta(G) +1$ colors
when $\Delta(G)\geq 3$. For some classes of perfect graphs, such as
triangulated graphs and comparability graphs, there are polynomial
algorithms to color the graph $G$ by $\chi(G)$ colors for
traditional coloring in \cite{Golumbic}. For these kinds of graphs,
one can also design polynomial algorithms to get the conditional
coloring number and the way to color these graphs, with a little
change of the original algorithms in \cite{Golumbic}. The details
are omitted.

\end{document}